\documentclass[11pt]{amsart}
\usepackage{fullpage}
\usepackage{amsmath, amssymb, amsthm,mathrsfs}
\usepackage{enumerate}
\usepackage{algorithm}
\usepackage {algpseudocode}
\usepackage{graphicx}
\usepackage{color}
\usepackage{bm}
\usepackage{bigints}
\usepackage{textgreek}
\usepackage[colorlinks=true,linkcolor=red,citecolor=blue,urlcolor=cyan]{hyperref}
\usepackage{verbatim}
\usepackage{enumitem}

\theoremstyle{remark}

\makeatletter
\newcommand\xleftrightarrow[2][]{%
  \ext@arrow 9999{\longleftrightarrowfill@}{#1}{#2}}
\newcommand\longleftrightarrowfill@{%
  \arrowfill@\leftarrow\relbar\rightarrow}
\makeatother

\usepackage[backend=bibtex,giveninits=true,sorting=nyt,natbib=true,maxcitenames=2,maxbibnames=10,url=false,doi=true,backref=false]{biblatex}

\renewbibmacro{in:}{\ifentrytype{article}{}{\printtext{\bibstring{in}\intitlepunct}}}

\makeatletter
\DeclareFontFamily{U}{tipa}{}
\DeclareFontShape{U}{tipa}{m}{n}{<->tipa10}{}
\newcommand{\arc@char}{{\usefont{U}{tipa}{m}{n}\symbol{62}}}%

\newcommand{\arc}[1]{\mathpalette\arc@arc{#1}}

\newcommand{\arc@arc}[2]{%
  \sbox0{$\m@th#1#2$}%
  \vbox{
    \hbox{\resizebox{\wd0}{\height}{\arc@char}}
    \nointerlineskip
    \box0
  }%
}
\makeatother

\usepackage{pgf}

\begin{document}
\title{Stochastic Gene Expression Model of Nuclear-to-Cell Ratio Homeostasis}

\author{Xuesong Bai, Thomas G. Fai}
\address{Department of Mathematics, Brandeis University, Waltham, MA}

\keywords{Nuclear-to-cell ratio, stochastic gene expression, chemical master equation, organelle size control}
\date{\today}

\begin{abstract} 
Cell size varies between different cell types, and between different growth and osmotic conditions. However, the nuclear-to-cell volume ratio (N/C ratio) remains nearly constant. In this paper, we build on existing deterministic models of N/C ratio homeostasis and develop a simplified protein translation model to study the effect of stochasticity on the N/C ratio homeostasis. We solve the corresponding chemical master equation and obtain the mean and variance of the N/C ratio. We also use a Taylor expansion approximation to study the effects of the system size on the fluctuations of the N/C ratio. We then combine the translation model with a cell division model to study the effects of extrinsic noises from cell division on the N/C ratio. Our model demonstrates that the N/C ratio homeostasis is maintained when the stochasticity in cell growth is taken into account, that the N/C ratio is largely determined by the gene fraction of nuclear proteins, and that the fluctuations in the N/C ratio diminish as the system size increases.
\end{abstract}

\maketitle

\section{Introduction and background}

\subsection{Introduction}

Cell size varies among different cell types, and within the same cell type, cell size may vary significantly under different growth and osmotic conditions. However, it has long been observed that the size of the nucleus scales with cell size so that the nuclear-to-cytoplasmic volume ratio (N/C ratio for short) remains nearly constant \cite{conklin1914cell,moore2019determination,neumann2007nuclear}. The N/C ratio has been hypothesized to regulate developmental timing such as the zygotic genome activation in early embryo development \cite{syed2021nuclear}, and may be crucial in the origin of animal multicellularity \cite{colgren2023evolution,olivetta2023nuclear}. Because the N/C ratio is maintained when the number of genomes within a cell (ploidy) is manipulated \cite{balachandra2022nuclear}, it is unlikely that the N/C ratio is determined by DNA content, and the mechanism for the determination of the N/C ratio remains an active field of study.

A simple osmotic pressure balance model was proposed in \cite{biswas2023conserved,deviri2022balance,lemiere2022control} that provides a physical basis for the N/C ratio. The model is based on the balance of osmotic pressure at the nuclear envelope and the cell membrane, and it shows that the N/C ratio is effectively determined by the ratio between the osmolyte numbers in the nucleus and the cytoplasm. However, the cellular mechanism that makes this osmolyte number ratio insensitive to perturbations is still poorly understood \cite{rollin2022cell}.

During cell growth, an N/C ratio homeostasis is maintained dynamically, where aberrant N/C ratios are corrected over time. A macroscopic growth model based on an active feedback mechanism was proposed in \cite{cantwell2019homeostatic}, whereas our previous work \cite{lemiere2022control} showed that the N/C ratio homeostasis can be explained by a growth model without active feedback. More detailed growth models that take into account the nucleocytoplasmic transport have also been proposed \cite{biswas2023conserved,wu2022correlation}. To the best of our knowledge, all these previous growth models explaining the N/C ratio homeostasis are deterministic.

In general, the dynamics in cellular processes are inherently stochastic, and mathematical models of such processes must incorporate stochasticity in order to effectively describe experimental observations \cite{wilkinson2009stochastic}. Indeed, we have observed fluctuations in the N/C ratio in previous experiments (Figure 7C in \cite{lemiere2022control}). Given that the process of cell growth, including mRNA transcription and protein translation, are stochastic in nature \cite{thomas2018sources}, here we explore a model of the fluctuations in the N/C ratio that arise from the inherently discrete nature of the transcription and translation processes.

In this paper, we develop a stochastic cell growth model to study the N/C ratio homeostasis, which is based on an existing stochastic gene expression model \cite{lin2018homeostasis}. We solve the corresponding chemical master equation (CME) to obtain a time-dependent joint probability distribution of the random variables. We then use this probability distribution to calculate the mean and variance of the N/C ratio and show the N/C ratio homeostasis is maintained in this stochastic model. Finally, we use a Taylor expansion approximation of the N/C ratio to study how the system size affects the fluctuations of the N/C ratio, and show that the fluctuations are predicted to diminish as the system size grow.

The major challenge faced in this study is solving the CME. To the best of our knowledge, while it is possible to analytically solve the CMEs corresponding to general systems of first order (monomolecular) reactions \cite{jahnke2007solving,reis2018general,vastola2021solving}, for second order (bimolecular) reactions and beyond, analytical solutions of the CMEs corresponding to general reaction systems are unavailable except for certain special cases \cite{laurenzi2000analytical,lee2012analytical}. Various numerical methods are available to compute approximate solutions, such as the Gillespie algorithm \cite{gillespie1976general,gillespie1977exact}, Galerkin spectral methods \cite{engblom2006discrete,engblom2009galerkin,engblom2009spectral}, and finite space projection methods \cite{munsky2006finite}. Analytical approximation methods for CMEs are also known, such as the chemical Langevin equations, the Fokker-Planck equations, and the system size expansion \cite{gardiner1985handbook,van1992stochastic}. 

In this paper, we use a minimal stochastic cell growth model in which all reactions are first order, which allows us to use a Fourier transform method \cite{reis2018general} to solve the CME. We achieve this relative simplicity by modeling the translation process only, instead of modeling both the transcription and the translation process. We use a coarse-grained model for the translation process that does not attempt to capture in detail the various cell biological processes involved in N/C ratio determination, such as nucleocytoplasmic transport. This minimal model allows us to capture salient aspects of N/C ratio homeostasis and provides a basis for more detailed stochastic models for the N/C ratio homeostasis in future studies. 

We will explain how in this paper we (i) incorporate stochasticity into models of nuclear size control by osmotic force balance; (ii) obtain an expression for the dependence of the N/C ratio on system size in the limit of a large number of solutes.

\subsection{Osmotic pressure balance model}\label{OsmoticBalance}

Our starting point is the osmotic pressure balance model \cite{biswas2023conserved,deviri2022balance,lemiere2022control}, where the  osmotic pressure is generated by chemical potential differences resulting from concentration differences in the solutes across semipermeable membranes \cite{wu2015simulation}. The basic assumptions of the osmotic pressure balance model are as follows:
\begin{enumerate}
 \item The osmotic effect of chromatin is negligible;
 \item The osmotic pressure is described by van't Hoff's law \cite{van1901osmotic};
 \item The tension at the nuclear envelope and the cell membrane is described by the Young-Laplace law.
\end{enumerate}
The equations describing the osmotic pressure balance are
\begin{subequations}\label{osmbal}
 \begin{align}
 (c_n-c_{cyto})k_BT &= \frac{2\sigma_n}{R_n}; \\
 (c_{cyto}-c_0)k_BT &= \frac{2\sigma_{cell}}{R_{cell}},
 \end{align}
\end{subequations}
in which $c_n$, $c_{cyto}$, and $c_0$ are the total concentrations of solutes in the nucleus, the cytoplasm, and outside the cell; $\sigma_n$ and $\sigma_{cell}$ are the tensions in the nuclear envelope and the cell membrane; $R_n$ and $R_{cell}$ are the radii of the nucleus and the cell, respectively; $k_B$ is the Boltzmann constant; and $T$ is the temperature. In 3D, $\kappa_n=1/R_n$ and $\kappa_{cell}=1/R_{cell}$ are the curvatures of the spheres representing the nucleus and cell, respectively. Hence, the solution of (\ref{osmbal}) represents a balance of osmotic forces with the elastic restoring forces $2\sigma_n\kappa_n$ and $2\sigma_{cell}\kappa_{cell}$ generated by surface tension.

If the further assumption is made that smaller osmolytes (e.g. ions and metabolites such as amino acids) freely diffuse across the nuclear envelope through nuclear pore complexes (NPCs), then only the concentrations of larger osmolytes (proteins and other macromolecules) are needed in the pressure balance of the nuclear envelope:
\begin{subequations}
\begin{align}
 \left(\frac{p_n}{V_n-b_n}-\frac{p_{cyto}}{V_{cyto}-b_{cyto}}\right)k_BT &= \frac{2\sigma_n}{R_n}; \\
 \left(\frac{p_{cyto}+n_{cyto}}{V_{cyto}-b_{cyto}}-c_0\right)k_BT &= \frac{2\sigma_{cell}}{R_{cell}}.
\end{align}
\end{subequations}
Here $n_{cyto}$ is the number of smaller osmolytes in the cytoplasm, $p$ denotes the number of the larger osmolytes in the respective compartment, $V$ denotes the total volume, and $b$ denotes the non-osmotic volume. We note that we use $p$ to denote solute number throughout this paper. In particular, it does not denote the osmotic pressure. 

In the case that the nuclear envelope is stress-free and $\sigma_n=0$, and that the non-osmotic volumes are negligible i.e. $b_n\ll V_n$ and $b_{cyto}\ll V_{cyto}$, the osmotic pressure balance equations simplify further \cite{lemiere2022control} to
\begin{equation}\label{NCyDef1}
 \frac{p_n}{V_n}=\frac{p_{cyto}}{V_{cyto}},\quad\mbox{i.e.}\quad \frac{V_n}{V_{cyto}}=\frac{p_n}{p_{cyto}}:=\Phi_{NCyto}.
\end{equation}
Equation (\ref{NCyDef1}) implies that under the assumptions stated above, by van't Hoff's formula, the N/C ratio is determined solely by the ratio of protein numbers in the nucleus and the cytoplasm. Therefore, we may study the N/C ratio homeostasis problem by studying the mechanisms that maintains the homeostasis of the protein ratio.

\subsection{A statistical viewpoint on the N/C ratio: stochastic protein translation models}

We next study the N/C ratio homeostasis problem from a statistical viewpoint. More precisely, we formulate a stochastic model for the growth of protein numbers which not only captures the passive feedback mechanism that corrects aberrant N/C ratio values \cite{lemiere2022control}, but also explains the fluctuations in the N/C ratio over time.

Consider a biochemical reaction system representing a simplified process of gene expression, with the species of reactants listed below.
\begin{itemize}
 \item Proteins: RNA polymerases (RNAPs) $P_p$, ribosomes $P_r$, nuclear proteins $P_n$, cytoplasmic proteins $P_c$;
 \item mRNAs: $M_p$, $M_r$, $M_n$, $M_c$, which corresponds to the above protein species;
 \item Genes: $G_p$, $G_r$, $G_n$, $G_c$, which corresponds to the above mRNA species.
\end{itemize}
Let $g_i$, $m_i$, $p_i$, $i=p,r,n,c$ denote the corresponding numbers of genes, mRNAs, and proteins, respectively. 

We begin by considering the stochastic gene expression model of Lin and Amir \cite{lin2018homeostasis}, which assumes that the transcription (resp. translation) rate for a specific gene $i$ is proportional to the fraction of RNAPs (resp. ribosomes) bound to its genes (resp. mRNAs), $\phi_i = \dfrac{g_i}{\sum g_i}$ (resp. $f_i = \dfrac{m_i}{\sum m_i}$):
\begin{subequations}\label{LinAmirModel}
\begin{align}
 \emptyset &\xrightarrow{k_0\frac{g_i}{\sum g_i}p_p}M_i,\quad i = p,r,n,c \quad\mbox{(transcription);} \\
 \emptyset &\xrightarrow{k_t\frac{m_i}{\sum m_i}p_r}P_i,\quad i = p,r,n,c \quad\mbox{(translation);} \\
 M_i &\xrightarrow{\frac{1}{\tau}}\emptyset,\quad i = p,r,n,c \quad\mbox{(mRNA degredation).}
\end{align}
\end{subequations}

\subsection{A simplified protein translation model}

We seek to obtain a complete time-dependent joint probability distribution of the random variables in the model. However, in the above transcription and translation model, the reaction rates are time-dependent, which makes the model difficult to solve analytically. Therefore, we consider the following simplified model of translation, which assumes that the translation rate for each protein is directly proportional to its gene fraction:
\begin{subequations}\label{SimplifiedModel}
\begin{align}
 P_r&\xrightarrow{k_r}2P_r, \quad k_r = k_t\phi_r \quad\mbox{(autocatalytic synthesis of ribosomes)};\\
 P_r&\xrightarrow{k_n}P_r+P_n, \quad k_n = k_t\phi_n \quad\mbox{(translation of nuclear proteins)};\\
 P_r&\xrightarrow{k_c}P_r+P_c, \quad k_c = k_t\phi_c \quad\mbox{(translation of cytoplasmic proteins)}.
\end{align}
\end{subequations}

Comparing the simplified model (\ref{SimplifiedModel}) with the original model (\ref{LinAmirModel}), we see that to assume the translation rate $k_i=k_t\phi_i$ means to assume that the mRNA fractions $f_i$ are equal to their corresponding gene fractions $\phi_i$. We intentionally choose this constant translation rate model because the reaction system involves only first-order reactions with time-independent reaction rates, so that its CME can be solved analytically. We use this simplified model (\ref{SimplifiedModel}) in all our stochastic simulations in this paper. The means of protein numbers (\ref{meanpr},\ref{meanpnpc}) calculated from the simplified model (\ref{SimplifiedModel}) are consistent with the corresponding results from the original model in \cite{lin2018homeostasis}. It may be interesting to compare the probability distributions of the protein numbers obtained from the original model and the simplified model, but such comparison goes beyond the scope of this paper.

\subsection{Nuclear-to-cytoplasmic ratio and nuclear-to-cell ratio}

We first note that in the literature, the N/C ratio usually denotes a ratio between volumes. However, as we have seen from (\ref{NCyDef1}), under the assumptions of our osmotic pressure balance model, the ratio between volumes is equal to the ratio of the corresponding protein numbers. Therefore, from this section on, we will focus on the ratio between protein numbers unless otherwise specified.

We note further that the N/C ratio may either refer in the literature to the nuclear-to-cell ratio ($\Phi_{NCell}$) or the nuclear-to-cytoplasmic ratio ($\Phi_{NCyto}$). To clarify, we define $\Phi_{NCyto}$ to be the ratio between the number of nuclear proteins and the total number of cytoplasmic proteins, as in (\ref{NCyDef1}):
\begin{equation*}
 \Phi_{NCyto} = \frac{p_n}{p_r+p_c},
\end{equation*} 
and we define $\Phi_{NCell}$ to be the ratio between the number of nuclear proteins and the total number of proteins in the cell:
\begin{equation*}
 \Phi_{NCell} = \frac{p_n}{p_r+p_n+p_c}.
\end{equation*}
We will focus on $\Phi_{NCell}$ here as in our previous work \cite{lemiere2022control}, although by (\ref{NCyDef1}) it may be more convenient in some cases to use $\Phi_{NCyto}$, e.g. when discussing the osmotic balance across the nuclear envelope.

The main results to be shown in this paper are as follows:
\begin{enumerate}
 \item We show that $\Phi_{NCell}$ is determined by the ratios of gene fractions $\frac{\phi_n}{\phi_r+\phi_n+\phi_c}=\phi_n$, where $\phi_i = \frac{g_i}{\sum g_i}$, and $\phi_r+\phi_n+\phi_c=1$;
  \item We show that $\Phi_{NCell}$ will maintain a homeostasis; namely, if the initial conditions
  \begin{equation*}
  \Phi_{NCyto,0} = \frac{p_{n,0}}{p_{r,0}+p_{n,0}+p_{c,0}}
  \end{equation*}
  deviate from $\phi_n$, then $\Phi_{NCell}\to\phi_n$ as time $t\to\infty$;
 \item We compute the mean, variance, and standard deviation of $\Phi_{NCell}$; in particular, we show how the coefficient of variation $\frac{\mathrm{SD}(\Phi_{NCell})}{\langle \Phi_{NCell} \rangle}$ decreases as the numbers of proteins $p_r$, $p_n$, and $p_c$ increase;
 \item We determine the relationship between
 \begin{equation*}
 \langle \Phi_{NCell} \rangle = \left\langle \frac{p_n}{p_r+p_n+p_c} \right\rangle\quad\mbox{and}\quad\frac{\langle p_n \rangle}{\langle p_r \rangle+\langle p_n \rangle+\langle p_c \rangle},
 \end{equation*}
  where the former is the average of the ratio and the latter is the ratio of the average.
\end{enumerate}

\section{Methods}

\subsection{Chemical master equation corresponding to the simplified translation model}

The CME corresponding to the reaction system (\ref{SimplifiedModel}) is
\begin{equation}\label{CME}\begin{split}
 &\frac{\partial}{\partial t}P(p_r,p_n,p_c,t) \\
 &= k_r(p_r-1)P(p_r-1,p_n,p_c,t) + k_np_rP(p_r,p_n-1,p_c,t) + k_cp_rP(p_r,p_n,p_c-1) \\
 &- (k_r+k_n+k_c)p_rP(p_r,p_n,p_c,t),
\end{split}\end{equation}
where $P(p_r,p_n,p_c,t)$ is the joint probability distribution of $p_r$, $p_n$, and $p_c$. 

The CME (\ref{CME}) can be solved with different initial conditions. A straightforward choice is the deterministic initial condition: the numbers of proteins at $t=0$ are given deterministically by $p_{r,0}$, $p_{n,0}$, and $p_{c,0}$, and
\begin{equation}
 P(p_r,p_n,p_c,0) = \delta_{p_r-p_{r,0}}\delta_{p_n-p_{n,0}}\delta_{p_c-p_{c,0}}.
\end{equation}
Alternatively, the initial condition can be stochastic instead of deterministic. From a biological view point, stochastic initial conditions can be used to model the random partitioning of molecules at cell division. An example is the Poisson initial condition: the averages of protein numbers at $t=0$ are given by $\langle p_r(0)\rangle = p_{r,0}$, $\langle p_n(0)\rangle = p_{n,0}$, and $\langle p_c(0)\rangle = p_{c,0}$; if $p_r(0)$, $p_n(0)$, and $p_c(0)$ are assumed to be independently Poisson distributed, then
\begin{equation}
 P(p_r,p_n,p_c,0) = \frac{p_{r,0}^{p_r}p_{n,0}^{p_n}p_{c,0}^{p_c}}{p_r!p_n!p_c!}e^{-(p_{r,0}+p_{n,0}+p_{c,0})}.
\end{equation}
Other types of initial conditions, such as binomial initial condition, can also be applied.

In what follows, we will only consider deterministic initial conditions, and we will specify initial conditions for (\ref{CME}) with emphasis on the size of the reaction system, as well as the initial $\Phi_{NCyto}$ and $\Phi_{NCell}$ ratios. To do this, we introduce a \textit{system size variable} $N$, as well as \textit{initial protein fractions} $f_r$, $f_n$, $f_c$, corresponding to ribosomes, nuclear, and cytoplasmic proteins, respectively, so that the initial $\Phi_{NCyto}$ and $\Phi_{NCell}$ ratios are
\begin{equation*}
NC_{cyto,0} = \frac{f_n}{f_r+f_c}, \quad \Phi_{NCyto,0} = \frac{f_n}{f_r+f_n+f_c}.
\end{equation*}
(Note here that we do not require $f_r+f_n+f_c=1$.) Then $p_{r,0}$, $p_{n,0}$, and $p_{c,0}$ are given by
\begin{equation*}
 p_{r,0}=f_r N,\quad p_{n,0}=f_n N,\quad p_{c,0}=f_c N.
\end{equation*}

\subsubsection{Solving the chemical master equation using the generating function and Fourier transform}

The generating function for the joint distribution $P(p_r,p_n,p_c,t)$ is defined as
\begin{equation}\label{generatingfunc}
 G(s_r,s_n,s_c,t) = \sum_{p_r=0}^\infty\sum_{p_n=0}^\infty\sum_{p_c=0}^\infty P(p_r,p_n,p_c,t)s_r^{p_r}s_n^{p_n}s_c^{p_c}.
\end{equation}

We will solve (\ref{CME}) by first solving the partial differential equation (PDE) for the generating function. Using a well-known formula for generating function equations \cite{gardiner1985handbook}, we get the PDE for $G(s_r,s_n,s_c,t)$:
\begin{equation}\label{generatingeqn}
 \frac{\partial}{\partial t}G(s_r,s_n,s_c,t) = [k_r(s_r^2-s_r)+k_n(s_rs_n-s_r)+k_c(s_rs_c-s_r)]\frac{\partial}{\partial s_r}G(s_r,s_n,s_c,t),
\end{equation}
with deterministic initial condition
\begin{equation}
 G(s_r,s_n,s_c,0) = s_r^{p_{r,0}}s_n^{p_{n,0}}s_c^{p_{c,0}}.
\end{equation}
For general reaction systems involving second or higher order reactions, the generating function equations are PDEs of at least second order, which, in general, cannot be solved analytically. However, since our reaction system (\ref{SimplifiedModel}) involves only first order reactions, the generating function equation (\ref{generatingeqn}) is a first order PDE and can thus be solved using the method of characteristics: the characteristic equations for (\ref{generatingeqn}) are
\begin{subequations}
\begin{align}
 \frac{dt}{d\tau} &= 1, \label{t} \\
 \frac{ds_r}{d\tau} &= -[k_r(s_r^2-s_r)+k_n(s_rs_n-s_r)+k_c(s_rs_c-s_r)], \label{sr} \\
 \frac{ds_n}{d\tau} &= 0, \label{sn} \\
 \frac{ds_c}{d\tau} &= 0, \label{sc} \\
 \frac{dG}{d\tau} &= 0. \label{G}
\end{align}
\end{subequations}
(\ref{t}), (\ref{sn}), (\ref{sc}), and (\ref{G}) lead to the following results:
\begin{subequations}
\begin{align}
 t &= \tau, \\
 s_n &= s_{n,0}, \\
 s_c &= s_{c,0}, \\
 G &= s_{r,0}^{p_{r,0}}s_{n,0}^{p_{n,0}}s_{c,0}^{p_{c,0}}. \quad\mbox{(Deterministic I.C.)} 
\end{align}
\end{subequations}
Now (\ref{sr}) becomes
\begin{equation}
 \frac{ds_r}{dt} = -[k_r(s_r^2-s_r)+k_n(s_rs_{n,0}-s_r)+k_c(s_rs_{c,0}-s_r)],
\end{equation}
which is separable and can be solved by the method of partial fractions. The solution is given by
\begin{equation}
 s_r = \frac{-k_rs_{r,0}+k_n(s_{n,0}-1)s_{r,0}+k_c(s_{c,0}-1)s_{r,0}}{[k_r(s_{r,0}-1)+k_n(s_{n,0}-1)+k_c(s_{c,0}-1)]e^{[-k_r+k_n(s_{n,0}-1)+k_c(s_{c,0}-1)]t}-k_rs_{r,0}},
\end{equation}
and then
\begin{equation}
 s_{r,0} = \frac{-k_rs_r+k_n(s_{n,0}-1)s_r+k_c(s_{c,0}-1)s_r}{[k_r(s_r-1)+k_n(s_{n,0}-1)+k_c(s_{c,0}-1)]e^{[k_r-k_n(s_{n,0}-1)-k_c(s_{c,0}-1)]t}-k_rs_r},
\end{equation}
Putting everything together, we get an expression of the generating function corresponding to deterministic initial condition
\begin{equation}\label{GeneratingD}\begin{split}
 &G(s_r,s_n,s_c,t) = \\
 &\left[\frac{-k_rs_r+k_n(s_n-1)s_r+k_c(s_c-1)s_r}{[k_r(s_r-1)+k_n(s_n-1)+k_c(s_c-1)]e^{[k_r-k_n(s_n-1)-k_c(s_c-1)]t}-k_rs_r}\right]^{p_{r,0}}s_n^{p_{n,0}}s_c^{p_c,0}.
\end{split}\end{equation}
Now we can calculate $P(p_r,p_n,p_c,t)$ by using the following Cauchy integral formula \cite{reis2018general}:
\begin{equation}\label{Cauchy}
 P(p_r,p_n,p_c,t) = \frac{1}{(2\pi i)^3}\oint_{C_r\times C_n \times C_c}\frac{G(s_r,s_n,s_c,t)}{s_r^{p_r+1}s_n^{p_n+1}s_c^{p_c+1}}ds_rds_nds_c,
\end{equation}
where $C_i$ are discs encircling the origin of the complex hyperplane. If we let $\psi(\theta_r,\theta_n,\theta_c,t) = G(e^{2\pi i\theta_r},e^{2\pi i\theta_n},e^{2\pi i\theta_c},t)$, then (\ref{Cauchy}) becomes
\begin{equation}\label{Fourier}
 P(p_r,p_n,p_c,t) = \int_{[0,1]^3}\psi(\theta_r,\theta_n,\theta_c,t)e^{-2\pi i\theta_rp_r}e^{-2\pi i\theta_np_n}e^{-2\pi i\theta_cp_c}d\theta_rd\theta_nd\theta_c,
\end{equation}
which is the Fourier transform of $\psi(\theta_r,\theta_n,\theta_c,t)$. Therefore, the analytical solution (\ref{Fourier}) of the CME (\ref{CME}) is given in terms of an integral rather than a closed-form solution. However, (\ref{Fourier}) can be conveniently evaluated numerically using, e.g. Matlab's \texttt{fftn}. As we will see later in Section \ref{meanexpgrowth}, the means of protein numbers grow exponentially. Therefore, the solution (\ref{Fourier}) is time-dependent and does not approach a stationary distribution as $t\to\infty$.

We note that it may be possible to obtain a closed-form solution of the CME (\ref{CME}) using random time change representation of Poisson processes \cite{anderson2011continuous,anderson2015stochastic}. However, such methods are beyond the scope of this paper.

\subsubsection{Calculation of moments from the generating function}

As a consequence of solving the CME (\ref{CME}) using the generating function and Fourier transform, the moments of the joint distribution can be conveniently calculated from the generating function (\ref{generatingfunc}) by taking the corresponding partial derivatives of the expression (\ref{GeneratingD}). These moments will later be used in the determination of grid sizes. For the first order moments (means) we have
\begin{equation}\label{genfunc1storder}
 \left\langle p_j(t) \right\rangle = \frac{\partial}{\partial s_j}G(s_r,s_n,s_c,t)\Big|_{s_r=s_n=s_c=1},\quad j=r,n,c.
\end{equation}
As for the second order moments, we can calculate the variances by
\begin{equation}
 \left\langle (p_j(t))(p_j(t)-1) \right\rangle = \frac{\partial^2}{\partial s_j^2}G(s_r,s_n,s_c,t)\Big|_{s_r=s_n=s_c=1},\quad j=r,n,c
\end{equation}
so that
\begin{equation}\label{genfunc2ndorder1}
 \left\langle p_j^2(t) \right\rangle = \frac{\partial^2}{\partial s_j^2}G(s_r,s_n,s_c,t)\Big|_{s_r=s_n=s_c=1}+\left\langle p_j(t) \right\rangle,\quad j=r,n,c;
\end{equation}
and for the covariances we have
\begin{equation}\label{genfunc2ndorder2}
 \left\langle p_j(t)p_k(t) \right\rangle = \frac{\partial^2}{\partial s_j \partial s_k}G(s_r,s_n,s_c,t)\Big|_{s_r=s_n=s_c=1},\quad j,k=r,n,c.
\end{equation}
In principle, higher order moments can be calculated in a similar manner, although the expressions would be considerably more complicated.

\subsubsection{Memory space problem and variable grid size method}\label{MemorySpace}

A major challenge in the numerical implementation of (\ref{Fourier}) is that the solution process described above is memory intensive. Suppose that a uniform grid is used and that the number of grid points for each dimension in \texttt{fftn} is $N_g$. Then we would need a complex 3D array of dimension $N_g\times N_g\times N_g$ to represent the discretized $\psi(\theta_r,\theta_n,\theta_c,t)$, which would occupy $16N_g^3$ Bytes of memory space. Therefore, the memory usage grows rapidly when $N$ increases. For example, when $N_g=1,024$, the memory usage is 16GB, which still lies within the capability of most desktop or laptop computers; whereas when $N_g=2,048$, the memory usage increases to 128GB, which is beyond the capability of many desktop computers (or even the capability of some high performance computing clusters).

In order to make the memory allocation more efficient, we proposed a variable grid size based on Chebyshev's inequality which assigns different number of grid points for each dimension, and also make the grid numbers vary with time $t$ to account for the evolution of the probability distribution. That is, instead of a uniform grid, we allow the grid sizes in each dimension to be different and time-dependent, so that the dimension of the 3D array is $N_r(t)\times N_n(t) \times N_c(t)$, and
\begin{equation}
 N_j(t) = \mbox{\texttt{ceileven}}(\langle p_j(t) \rangle+N_{SD}\mathrm{SD}(p_j(t))),
\end{equation}
where the function \texttt{ceileven}$(u)$ means rounding $u$ to the nearest even integer greater than or equal to $u$, and $N_{SD}$ is the number of standard deviations away from the mean. Note that $N_j(t)$'s need to be even in order for proper implementation of \texttt{fftn}. By Chebyshev's inequality, taking $N_{SD}=8$ will yield a grid that covers at least 98\% of the distribution. The means $\langle p_j(t) \rangle$ and standard deviations $\mathrm{SD}(p_j(t))$ can be calculated from (\ref{genfunc1storder}) and (\ref{genfunc2ndorder1}) without knowing the full solution of the CME.

\subsubsection{Calculation of the mean and variance of the N/C ratio}

Once the joint distribution $P(p_r,p_n,p_c,t)$ is obtained, we can directly compute the first and second moment of $\Phi_{NCell}$ by
\begin{subequations}
\begin{align}
 \langle \Phi_{NCell}(t) \rangle &= \left\langle\frac{p_n(t)}{p_r(t)+p_n(t)+p_c(t)}\right\rangle = \sum_{\substack{p_r\geqslant 0, p_n\geqslant 0, p_c\geqslant 0 \\ p_r+p_n+p_c\neq 0}}\frac{p_n}{p_r+p_n+p_c}P(p_r,p_n,p_c,t), \\
 \langle \Phi_{NCell}(t)^2 \rangle &= \left\langle\left(\frac{p_n(t)}{p_r(t)+p_n(t)+p_c(t)}\right)^2\right\rangle = \sum_{\substack{p_r\geqslant 0, p_n\geqslant 0, p_c\geqslant 0 \\ p_r+p_n+p_c\neq 0}}\left(\frac{p_n}{p_r+p_n+p_c}\right)^2P(p_r,p_n,p_c,t). \\
\end{align}
\end{subequations}
The means are then equal to the first moments, while the variances are then calculated by
\begin{equation}
 \mathrm{Var}(\Phi_{NCell}(t)) = \langle \Phi_{NCell}^2(t) \rangle - \langle \Phi_{NCell}(t) \rangle^2.
\end{equation}

\subsection{Taylor expansion approximation of the N/C ratio}

It is possible to approximate the mean and variance of $\Phi_{NCell}$ and $\Phi_{NCyto}$ without calculating the joint probability distribution. This is done by using the bivariate Taylor expansion of the ratio function \cite{elandt1980survival}. A brief description of the method is given as follows.

Consider two general random variables $x_1$, $x_2$, where $x_2\neq 0$. Let $f(x_1,x_2)=\dfrac{x_1}{x_2}$ be the ratio of the two variables. The formal Taylor expansion $T(x_1,x_2)$ of $f(x_1,x_2)$ around the mean values $(\mu_1,\mu_2)=\left(\langle x_1 \rangle, \langle x_2 \rangle\right)$ is
\begin{equation}
 T(x_1,x_2) = \sum_{\alpha_1+\alpha_2\geqslant 0}\frac{\partial^{\alpha_1+\alpha_2}f(\mu_1,\mu_2)}{\partial x_1^{\alpha_1} \partial x_2^{\alpha_2}}\frac{(x_1-\mu_1)^{\alpha_1}(x_2-\mu_2)^{\alpha_2}}{\alpha_1!\alpha_2!}.
\end{equation}
Then the mean of the ratio $\left\langle \dfrac{x_1}{x_2} \right\rangle = \left\langle f(x_1,x_2) \right\rangle$ can be calculated formally by the taking mean of the Taylor expansion:
\begin{equation}\label{TaylorApprox1st}
 \left\langle f(x_1,x_2) \right\rangle = \left\langle T(x_1,x_2) \right\rangle = \sum_{\alpha_1+\alpha_2\geqslant 0}\frac{\partial^{\alpha_1+\alpha_2}f(\mu_1,\mu_2)}{\partial x_1^{\alpha_1} \partial x_2^{\alpha_2}} \frac{\left\langle(x_1-\mu_1)^{\alpha_1}(x_2-\mu_2)^{\alpha_2}\right\rangle}{\alpha_1!\alpha_2!},
\end{equation}
where $\left\langle(x_1-\mu_1)^{\alpha_1}(x_2-\mu_2)^{\alpha_2}\right\rangle$ are the mixed central moments of order $\alpha_1+\alpha_2$.

To obtain an approximation of the variance of the ratio $$\mathrm{Var}\left(\dfrac{x_1}{x_2}\right) = \mathrm{Var}(f(x_1,x_2)) = \left\langle f(x_1,x_2)^2 \right\rangle-\left\langle f(x_1,x_2) \right\rangle^2,$$ we need to first calculate formally $\left\langle f(x_1,x_2)^2 \right\rangle$ using the square of the formal Taylor expansion:
\begin{equation}
\begin{aligned}
 &T(x_1,x_2)^2 \\
 &= \sum_{\alpha_1+\alpha_2\geqslant 0}\sum_{\beta_1+\beta_2\geqslant 0}\frac{\partial^{\alpha_1+\alpha_2}f(\mu_1,\mu_2)}{\partial x_1^{\alpha_1} \partial x_2^{\alpha_2}}\frac{\partial^{\beta_1+\beta_2}f(\mu_1,\mu_2)}{\partial x_1^{\beta_1} \partial x_2^{\beta_2}}\frac{(x_1-\mu_1)^{\alpha_1+\beta_1}(x_2-\mu_2)^{\alpha_2+\beta_2}}{\alpha_1!\alpha_2!\beta_1!\beta_2!},
\end{aligned}
\end{equation}
so that
\begin{equation}\label{TaylorApprox2nd}\begin{split}
&\left\langle f(x_1,x_2)^2 \right\rangle = \left\langle T(x_1,x_2)^2 \right\rangle \\
&= \sum_{\alpha_1+\alpha_2\geqslant 0}\sum_{\beta_1+\beta_2\geqslant 0}\frac{\partial^{\alpha_1+\alpha_2}f(\mu_1,\mu_2)}{\partial x_1^{\alpha_1} \partial x_2^{\alpha_2}}\frac{\partial^{\beta_1+\beta_2}f(\mu_1,\mu_2)}{\partial x_1^{\beta_1} \partial x_2^{\beta_2}}\frac{\left\langle (x_1-\mu_1)^{\alpha_1+\beta_1}(x_2-\mu_2)^{\alpha_2+\beta_2}\right\rangle}{\alpha_1!\alpha_2!\beta_1!\beta_2!},
\end{split}\end{equation}
where, again, $\left\langle (x_1-\mu_1)^{\alpha_1+\beta_1}(x_2-\mu_2)^{\alpha_2+\beta_2}\right\rangle$ are the mixed central moments of order $\alpha_1+\alpha_2+\beta_1+\beta_2$. The variance is then given by
\begin{equation}\label{VarApprox}
 \mathrm{Var}(f(x_1,x_2)) = \mathrm{Var}(T(x_1,x_2)) = \left\langle T(x_1,x_2)^2 \right\rangle-\left\langle T(x_1,x_2) \right\rangle^2
\end{equation}

From (\ref{TaylorApprox1st}) and (\ref{TaylorApprox2nd}) we see that, in principle, as long as we can calculate (or approximately calculate)  all the mixed central moments $\left\langle (x_1-\mu_1)^{\gamma_1}(x_2-\mu_2)^{\gamma_2}\right\rangle$, we can then calculate Taylor expansion approximations of the mean and variance of the ratio. In practice, however, higher order moments tend to be difficult to calculate. Therefore, we will only implement Taylor expansion approximations of order up to 2. Later, we will see that for calculating approximations of the mean and variance of $\Phi_{NCell}$, these low order approximations work well compared to the more accurate results obtained from the joint probability distributions.

For the mean $\left\langle\dfrac{x_1}{x_2}\right\rangle$ of the ratio, it is feasible to calculate the Taylor expansion approximation up to the second order. However, we will start by looking at the first order approximation: letting $\alpha_1+\alpha_2\leqslant 1$ and substituting the partial derivatives of $f(x_1,x_2)=\dfrac{x_1}{x_2}$ into (\ref{TaylorApprox1st}), we get
\begin{equation}\label{mean1st}
 \left\langle\dfrac{x_1}{x_2}\right\rangle \approx \frac{\mu_1}{\mu_2} = \frac{\langle x_1 \rangle}{\langle x_2 \rangle}.
\end{equation}
This shows that \textit{the first order Taylor expansion approximation of the mean of the ratio is simply the ratio of the means}. A more accurate result can be achieved by the second order approximation: letting $\alpha_1+\alpha_2\leqslant 2$ and substituting the partial derivatives of $f(x_1,x_2)=\dfrac{x_1}{x_2}$ into (\ref{TaylorApprox1st}) again, we get
\begin{equation}\label{mean2nd}\begin{split}
 \left\langle\dfrac{x_1}{x_2}\right\rangle &\approx \frac{\mu_1}{\mu_2}-\frac{\left\langle (x_1-\mu_1)(x_2-\mu_2)\right\rangle}{\mu_2^2}+\frac{\left\langle (x_2-\mu_2)^2\right\rangle \mu_1}{\mu_2^3} \\
 &= \frac{\langle x_1 \rangle}{\langle x_2 \rangle}-\frac{\mathrm{Cov}(x_1,x_2)}{\langle x_2 \rangle^2}+\frac{\mathrm{Var}(x_2) \langle x_1 \rangle}{\langle x_2 \rangle^3}.
\end{split}\end{equation}

For the variance $\mathrm{Var}\left(\dfrac{x_1}{x_2}\right)$ of the ratio, it is only feasible to calculate the Taylor expansion approximation up to the first order, due to the complexity of the expression (\ref{TaylorApprox2nd}). In this case, we will also use the first order approximation (\ref{mean1st}) of the mean in the final calculation of the variance (\ref{VarApprox}). The expression of the first order Taylor expansion approximation of $\mathrm{Var}\left(\dfrac{x_1}{x_2}\right)$ is
\begin{equation}\begin{split}
 \mathrm{Var}\left(\dfrac{x_1}{x_2}\right) &\approx \frac{\left\langle (x_1-\mu_1)^2 \right\rangle}{\mu_2^2} - \frac{2\left\langle (x_1-\mu_1)(x_2-\mu_2) \right\rangle\mu_1}{\mu_2^3} + \frac{\left\langle (x_2-\mu_2)^2 \right\rangle\mu_1^2}{\mu_2^4} \\
 &= \frac{\mu_1^2}{\mu_2^2}\left[\frac{\left\langle (x_1-\mu_1)^2 \right\rangle}{\mu_1^2} - \frac{2\left\langle (x_1-\mu_1)(x_2-\mu_2) \right\rangle}{\mu_1\mu_2} + \frac{\left\langle (x_2-\mu_2)^2 \right\rangle}{\mu_2^2}\right] \\
 &= \frac{\langle x_1 \rangle^2}{\langle x_2 \rangle^2}\left[\frac{\mathrm{Var}(x_1)}{\langle x_1 \rangle^2}-\frac{2\mathrm{Cov}(x_1,x_2)}{\langle x_1 \rangle \langle x_2 \rangle}+\frac{\mathrm{Var}(x_2)}{\langle x_2 \rangle^2}\right].
\end{split}\end{equation}

\section{Results}

The Matlab codes for the following results are available at \href{https://github.com/topgunbai683/translation-nc-ratio.git}{github.com/topgunbai683/translation-nc-ratio.git}.

\subsection{Closed-form expression of moments}

As a by-product of the FFT method, we obtain closed-form expressions of the generating functions corresponding to the CME (\ref{CME}) with deterministic initial conditions (\ref{GeneratingD}), from which we can derive closed form expressions for the moments of the variables $p_r$, $p_n$, and $p_c$. These closed form expressions will play a key role later in the Taylor series expansion approximation of the $\Phi_{NCell}$ curves.

All the expressions of moments are calculated from symbolic differentiations of the generating function (\ref{GeneratingD}), using Matlab's Symbolic Toolbox.

It should be noted that these expressions can also be calculated by solving the ordinary differential equations (ODEs) for the moments, where the ODEs themselves are derived using the generating function equation (\ref{generatingeqn}) and (\ref{genfunc1storder}), (\ref{genfunc2ndorder1}), (\ref{genfunc2ndorder2}), or using the Kramers-Moyal expansion \cite{gardiner1985handbook,van1992stochastic}. 

\subsubsection{First order moments: means of protein numbers}\label{meanexpgrowth}

Using (\ref{genfunc1storder}), we can calculate the first order moments, that is, the means of each individual variable, as functions of  time. For the mean of ribosome numbers we have
\begin{equation}\label{meanpr}
 \langle p_r(t) \rangle = p_{r,0} {e}^{k_t \phi_r t} = f_r N {e}^{k_t \phi_r t},
\end{equation}
that is, the mean number of ribosomes grows exponentially with growth rate $k_r=k_t\phi_r$, which is expected from the autocatalytic replication of ribosomes. For the mean numbers of nuclear and cytoplasmic proteins we have
\begin{equation}\label{meanpnpc}
 \langle p_n(t) \rangle = p_{n,0}+p_{r,0} \left(\frac{\phi_n {e}^{k_t \phi_r t}}{\phi_r}-\frac{\phi_n}{\phi_r}\right) = f_n N+f_r N \left(\frac{\phi_n {e}^{k_t \phi_r t}}{\phi_r}-\frac{\phi_n}{\phi_r}\right),
\end{equation}
and
\begin{equation}
 \langle p_c(t) \rangle = p_{c,0}+p_{r,0} \left(\frac{\phi_c {e}^{k_t \phi_r t}}{\phi_r}-\frac{\phi_c}{\phi_r}\right)  = f_c N+f_r N \left(\frac{\phi_c {e}^{k_t \phi_r t}}{\phi_r}-\frac{\phi_c}{\phi_r}\right),
\end{equation}
respectively. This means that after some initial relaxation period, the mean numbers of nuclear and cytoplasmic proteins will also exhibit exponential growth with the same growth rate $k_r=k_t\phi_r$ as the growth of ribosomes. 

Overall, the expressions of the first moments exhibit an exponential growth of all the cellular proteins driven by the autocatalytic replication of ribosomes, which agrees with the previous model by Lin and Amir \cite{lin2018homeostasis}.

\subsubsection{Second order moments: variances and coefficients of variation}

Using (\ref{genfunc2ndorder1}) and (\ref{genfunc2ndorder2}), we can calculate the second order moments, and thereby the variances and covariances, of the variables. The expressions for the variances and covariances are technical and the formulas by themselves do not provide too much insight, so it is more useful to examine the coefficient of variation of each variable, which is defined as the ratio of the standard deviation and the mean, and measures the magnitude of fluctuations relative to the mean:
\begin{equation}
 \mathrm{CV}(p_j(t)) = \frac{\mathrm{SD}(p_j(t))}{\langle p_j(t) \rangle} = \frac{\sqrt{\mathrm{Var}(p_j(t))}}{\langle p_j(t) \rangle}, \quad j=r,n,c.
\end{equation}
For deterministic initial conditions, we have
\begin{equation}
 \mathrm{CV}(p_r(t)) = \frac{{e}^{-k_t \phi_r t} \sqrt{f_r {e}^{k_t \phi_r t} \left({e}^{k_t \phi_r t}-1\right)}}{f_r \sqrt{N}},
\end{equation}
\begin{equation}
 \mathrm{CV}(p_n(t)) = \frac{\sqrt{-f_r \phi_n \left(\phi_n+\phi_r-\phi_n {e}^{2 k_t \phi_r t}-\phi_r {e}^{k_t \phi_r t}+2 k_t \phi_n \phi_r t {e}^{k_t \phi_r t}\right)}}{\left(f_n \phi_r-f_r \phi_n+f_r \phi_n {e}^{k_t \phi_r t}\right) \sqrt{N}},
\end{equation}
and
\begin{equation}
 \mathrm{CV}(p_c(t)) = \frac{\sqrt{-f_r \phi_c \left(\phi_c+\phi_r-\phi_c {e}^{2 k_t \phi_r t}-\phi_r {e}^{k_t \phi_r t}+2 k_t \phi_c \phi_r t {e}^{k_t \phi_r t}\right)}}{\left(f_c \phi_r-f_r \phi_c+f_r \phi_c {e}^{k_t \phi_r t}\right) \sqrt{N}}.
\end{equation}
From the above equations, we see that the for each variable, the fluctuations relative to the mean are of order $N^{-\frac12}$. Therefore, if the system size is large, then the relative fluctuations become negligible.

It is known \cite{delbruck1940statistical} that, in the case of deterministic initial conditions, the ribosome number $p_r$ follows a negative binomial distributions, which is due to the autocatalytic replication of ribosomes. Thus, it will naturally have relative fluctuations of order $N^{-\frac12}$.

\subsection{Comparison of methods for solving chemical master equations: FFT vs SSA vs CLE}

Next we present the joint distribution $P(p_r,p_n,p_c,t)$ obtained by solving the chemical master equation (\ref{CME}), and give a comparison of the methods we used for solving (\ref{CME}). The main purpose of this comparison is to demonstrate a selection of methods that make it possible to calculate the joint distributions of the autocatalytic reaction system (\ref{SimplifiedModel}) for a wide enough range of timespans and system sizes. The calculated distributions will then be used to calculate $\Phi_{NCell}$ ratio curves. 

\begin{figure}[h]
 \includegraphics[width=\textwidth]{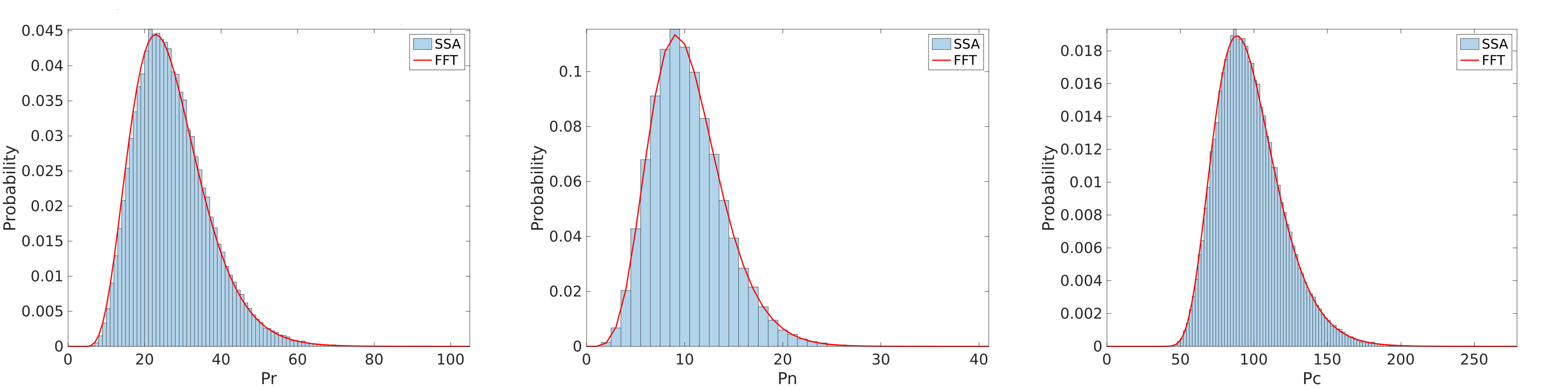}
 \caption{Comparison of marginal distributions calculated from FFT and Gillespie algorithm at $t=1500$. Left, ribosomes; middle, nuclear proteins; right, cytoplasmic proteins. System size $N = 30$.}
 \label{Marginal_dist_comparison_1500}
\end{figure}
\subsubsection{Gillespie algorithm}

The Gillespie algorithm, or stochastic simulation algorithm (SSA) \cite{gillespie1976general,gillespie1977exact}, is one of the most widely-used numerical methods for CMEs, and here the histograms generated by the SSA are used as a benchmark for evaluating the other methods.

The major advantage of the SSA is that it works for larger system sizes (N$\sim$ $10^5$ to $10^6$) and long time spans ($t\geqslant3000$s for $k_t = 0.005$). Although the simulations can be time consuming when $N$ is large, the SSA may be parallelized to accelerate the simulations.

However, instead of calculating the distribution functions, the SSA only samples the underlying distribution. To obtain a true numerical solution of the CME (\ref{CME}), we  therefore require other methods.

\subsubsection{FFT method}

Comparing the marginal distribution functions obtained from the FFT method and the histograms generated by the SSA (Fig. \ref{Marginal_dist_comparison_1500}), we see that the FFT method works reliably for the CME (\ref{CME}) with deterministic initial conditions. Therefore, when it is tractable to use, the FFT method is our preferred the method for calculating the joint distributions used for the $\Phi_{NCell}$ ratio curves.

However, due to the memory space limitation mentioned before, the FFT method only works for relatively small system sizes and short timespans. For example, on a computer with 16GB RAM, the FFT method will not run into memory space issues if the system size $N\sim 10^2$ and the time span $t\leqslant 3000$s. Although it would be possible to test the FFT method on computing clusters with much larger memory space, the aforementioned range of system size and timespan is already enough for the purpose of this paper.

\begin{figure}[h]
 \includegraphics[width=\textwidth]{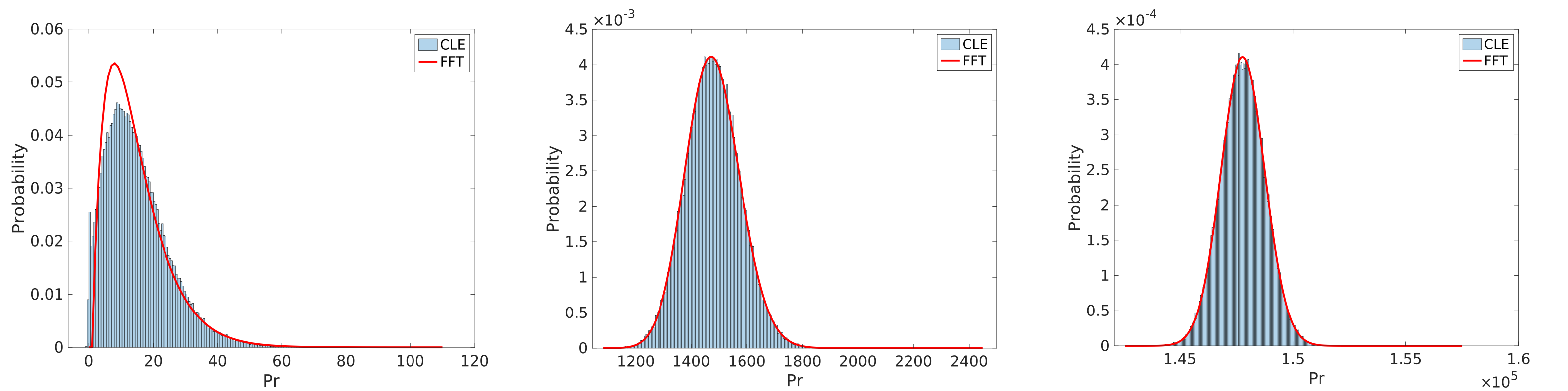}
 \caption{Comparison of marginal distributions of ribosome numbers calculated from FFT and CLE, with different system sizes. Left, $N = 10^1$; middle, $N=10^3$; right, $N=10^5$.}
 \label{CLE_FFT_comparison}
\end{figure}
\subsubsection{Chemical Langevin equation approximation}

We have shown in \ref{MemorySpace} that solving the CME (\ref{CME}) using the Fourier transform is a memory intensive process, especially when the system size variable $N$ increases and the number of grid points increases correspondingly. For biologically realistic values of $N$ with an order of magnitude of at least $10^4$, the Fourier transform method becomes numerically impractical, and approximation methods are needed to obtain the probability distributions from the CME (\ref{CME}). The chemical Langevin equation (CLE) is a widely-used method for large system size approximation, which can be derived by, e.g. the Kramers-Moyal expansion \cite{gardiner1985handbook,van1992stochastic}, or continuous time Markov chain models for chemical reactions \cite{anderson2011continuous,anderson2015stochastic}.

The CLEs corresponding to the simplified translation model (\ref{SimplifiedModel}) are given by
\begin{subequations}\label{CLE}
\begin{align}
 dp_r &= k_rp_rdt+\sqrt{k_rp_r}dW_r(t), \\
 dp_n &= k_np_rdt+\sqrt{k_np_r}dW_n(t), \\
 dp_c &= k_cp_rdt+\sqrt{k_cp_r}dW_c(t),
\end{align}
\end{subequations}
where $W_l(t)$ are independent standard Brownian motions, $l=r,n,c$. Now (\ref{CLE}) is a system of stochastic differential equations, which can be solved numerically using, e.g. Euler-Maruyama method \cite{higham2001algorithmic}. Then, the probability distributions can be obtained by generating a large enough number of sample paths from (\ref{CLE}).

Fig. \ref{CLE_FFT_comparison} shows a comparison of marginal distributions calculated from the FFT method and the CLE approximation. Here, we are using the marginal distribution of ribosome numbers as an example. We see that the CLE approximation does not work well when the system size variable $N$ is small, that is, when $N\sim 10^1$, and that the CLE approximation performs better as $N$ increases. When $N\sim 10^5$, the distributions calculated from FFT and CLE are nearly identical. Therefore, the CLE approximation is a valid method for simulations with large system sizes, e.g. simulations where $N\sim 10^7$ to $10^8$ for which the SSA becomes numerically impractical.

\subsection{N/C ratio curves from the FFT method}

Next we present the main results on the $\Phi_{NCell}$ ratio curves calculated from the FFT method with deterministic initial conditions. We assume that for a particular cell immediately after cell division, the initial condition for the number of intracellular proteins is deterministic. In other words, Poisson or other stochastic initial conditions will only be relevant in the case of an ensemble of cells. For the $\Phi_{NCell}$ ratio of a single cell, it therefore suffices to consider deterministic initial conditions.

\subsubsection{Maintaining homeostasis of the N/C ratio via growth}

\begin{figure}[h]
 \includegraphics[width=0.4\textwidth]{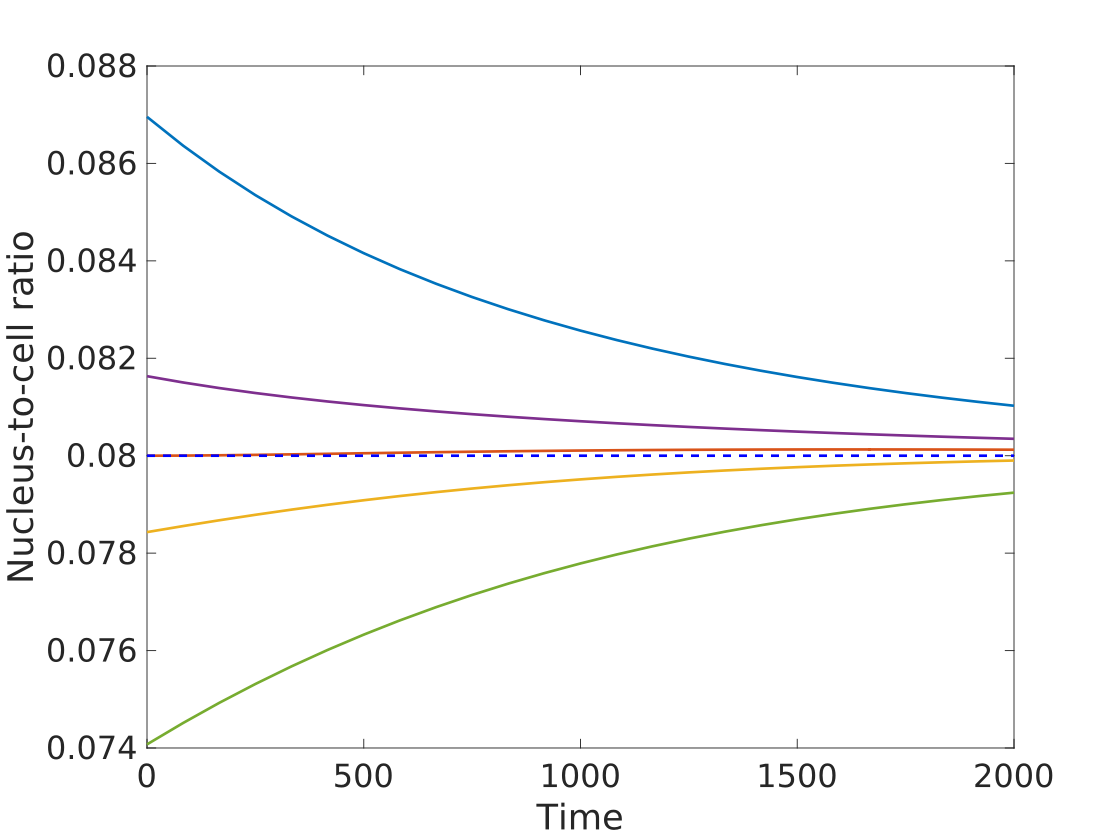}
 \caption{Mean nuclear-to-cell ratio curves with different initial conditions, showing relaxation to the steady-state value. System size $N = 100$.}
 \label{NCeMultipleCurves}
\end{figure}

Fig. \ref{NCeMultipleCurves} shows that the mean $\Phi_{NCell}$ ratio curves eventually approach $\phi_n=0.08$ if the initial ratio $\Phi_{NCyto,0}$ deviates from $\phi_n$. In other words, for this simplified translation model in which the protein numbers $p_r$, $p_n$, $p_c$ grow exponentially without bound, although the protein numbers themselves do not have steady states (other than the trivial steady state at 0), the ratio $\Phi_{NCell} = \dfrac{p_n}{p_r+p_n+p_c}$ does approach the steady state $\phi_n$.

Interestingly, if the initial N/C ratio $\Phi_{NCell,0}$ is exactly equal to $\phi_n$, then due to fluctuations, the mean $\Phi_{NCell}$ ratio curve may overshoot the steady state $\phi_n$. This overshot phenomenon diminishes as the system size variable $N$ increases, as will be shown later in the discussion of Taylor expansion approximation of the $\Phi_{NCell}$ ratio (See Fig. \ref{NCeSystemSizeComparisonTaylor} (b)).

This overshoot phenomenon can be partially explained by the stochastic differential equation (SDE) for $\Phi_{NCell}$. Using It\^{o}'s formula \cite{gardiner1985handbook}, we derive the following SDE from the CLEs (\ref{CLE}):
\begin{equation}\label{NCSDE}
\begin{aligned}
 d\Phi_{NCell} &= \left(\frac{k_tp_r}{p_r+p_n+p_c}-\frac{k_tp_r}{(p_r+p_n+p_c)^2}\right)(\phi_n-\Phi_{NCell})dt \\
 &+ \frac{(p_r+p_c)\sqrt{k_np_r}}{(p_r+p_n+p_c)^2}dW_n(t)-\frac{p_n\sqrt{k_rp_r}}{(p_r+p_n+p_c)^2}dW_r(t)-\frac{p_n\sqrt{k_cp_r}}{(p_r+p_n+p_c)^2}dW_c(t).
\end{aligned}
\end{equation}
Since the right-hand side of (\ref{NCSDE}) depends on other variables, we need to solve (\ref{NCSDE}) together with (\ref{CLE})\footnote{In fact, we only need two equations in (\ref{CLE}) to go together with (\ref{NCSDE}), because $\Phi_{NCell}$ depends on $p_r$, $p_n$, and $p_c$.}. However, we see from (\ref{NCSDE}) that even if the initial condition is $\Phi_{NCell,0}=\phi_n$, $\Phi_{NCell}$ will deviate from $\phi_n$ due to the noise terms.


Fig. \ref{NCeMultipleCurves} also shows that the more the initial ratio $\Phi_{NCyto,0}$ deviates from $\phi_n$, the faster the mean $\Phi_{NCell}$ ratio will return to $\phi_n$, in the sense that there is a larger absolute value of the rate of change at any given time. It is worth noting that this behavior is achieved without any active feedback mechanism in the model, in agreement with our previous deterministic growth model \cite{lemiere2022control}.

\subsubsection{Fluctuations of the N/C ratio and system size}
\begin{figure}[h]
 \includegraphics[width=\textwidth]{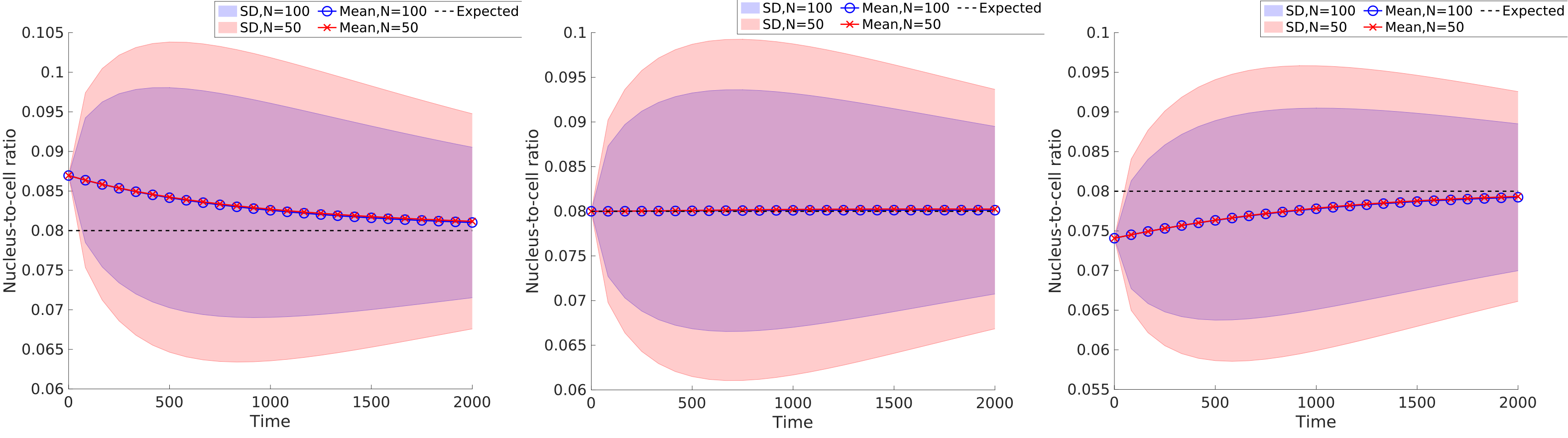}
 \caption{Nuclear-to-cell ratio curves with different initial conditions and system sizes. The three panels show results starting from different initial conditions.}
 \label{NCeSystemSizeComparison}
\end{figure}

Fig. \ref{NCeSystemSizeComparison} shows that the standard deviations of the $\Phi_{NCell}$ ratio curves decrease as the system size $N$ increases. We will study the how the system size affects the magnitude of fluctuations in the $\Phi_{NCell}$ ratio next in more detail.

\subsection{N/C ratio curves from Taylor expansion approximation}

Now we present the results on the $\Phi_{NCell}$ ratio curves calculated from the Taylor expansion approximation. 

\subsubsection{Validity of Taylor expansion approximation}
\begin{figure}[h]
 \includegraphics[width=\textwidth]{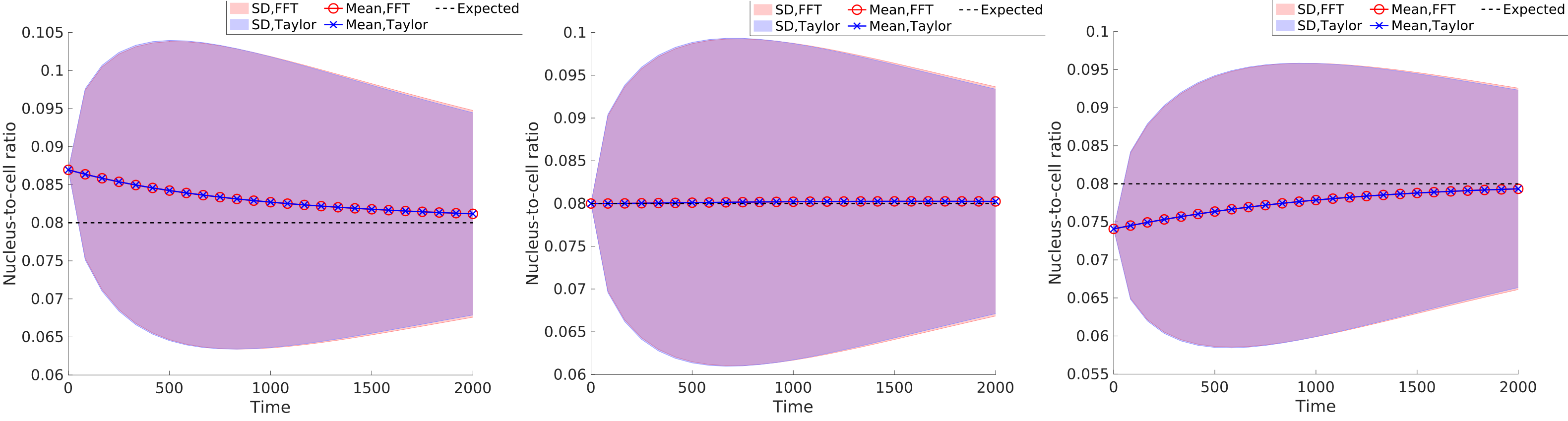}
 \caption{Nuclear-to-cell ratio curves calculated from FFT and Taylor expansion. The three panels show results results starting from different initial conditions.}
 \label{NCeTaylorFFTComparison_50}
\end{figure}

Fig. \ref{NCeTaylorFFTComparison_50} shows examples of mean $\Phi_{NCell}$ ratio curves and the corresponding standard deviations obtained from both the FFT method and the Taylor expansion approximation. We see that for a given initial condition, the mean $\Phi_{NCell}$ ratio curves obtained by the two methods are nearly identical, while the standard deviations show small but noticeable difference. This comparison demonstrates that the Taylor expansion approximation agrees with the $\Phi_{NCell}$ ratio calculated by the FFT method.

\subsubsection{Expressions of Taylor series expansion approximations}

Using the closed-form expressions of the means and variances of $p_r$, $p_n$, and $p_c$, we obtain explicit expressions of the Taylor series expansion approximation of the mean and variance of the $\Phi_{NCell}$ ratio. Because these expressions in their explicit forms are quite technical, we present the results in a more concise form, which conveys the important information in a clearer way.

\begin{figure}[h]
 \includegraphics[width=\textwidth]{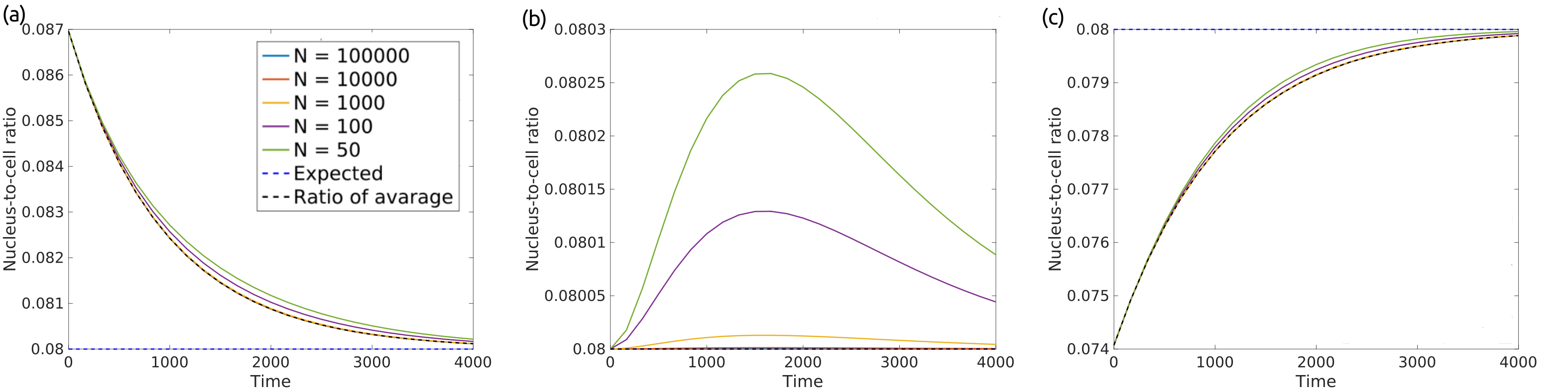}
 \caption{Nuclear-to-cell ratio curves with different initial conditions and system sizes, calculated from Taylor expansion. Note that all three panels share the same legend, which is shown in panel (a).}
 \label{NCeSystemSizeComparisonTaylor}
\end{figure}

\begin{figure}[h]
 \includegraphics[width=0.4\textwidth]{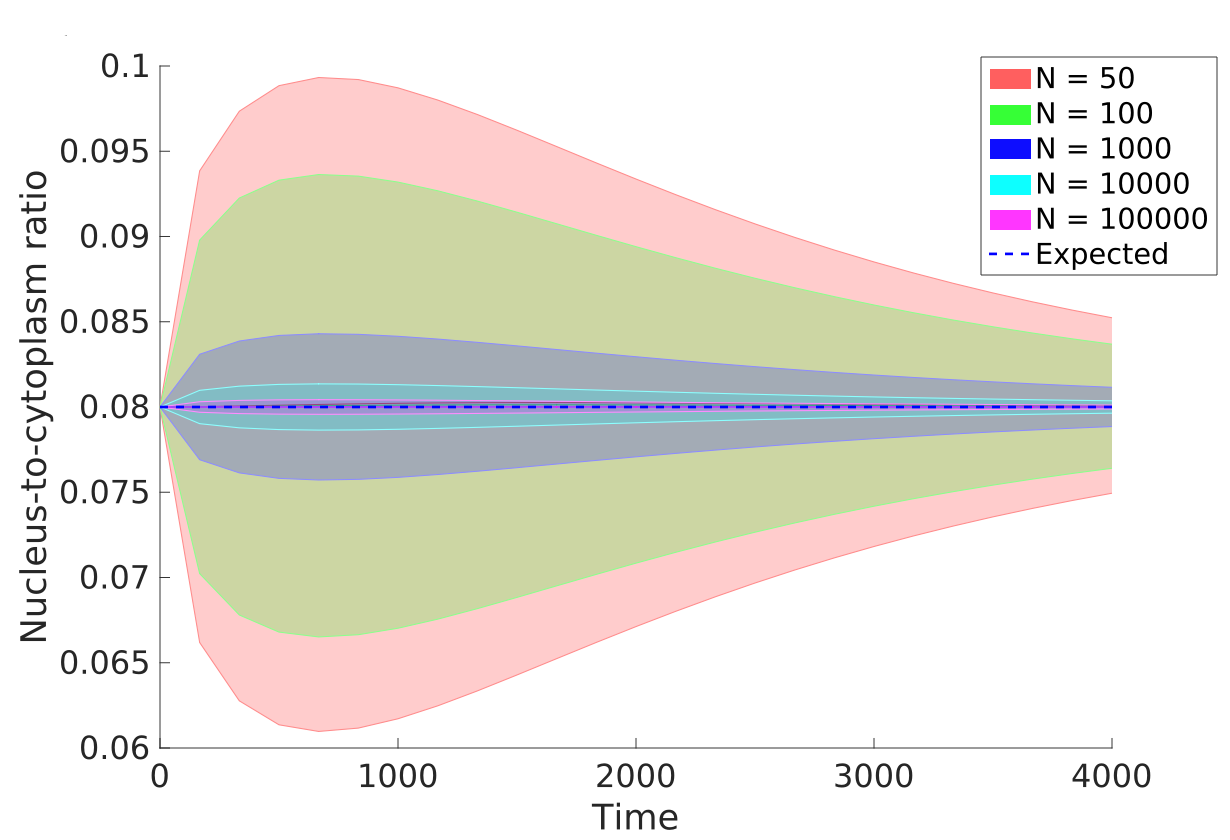}
 \caption{Standard deviation of the nuclear-to-cell ratio curve with different system sizes and initial condition $\phi_n=0.08$, calculated from Taylor expansion.}
 \label{NCeVarianceComparison}
\end{figure}

The expression for the mean of $\Phi_{NCell}$ has the form
\begin{equation}\label{ENCyTaylor}
\begin{aligned}
 \langle \Phi_{NCell} \rangle &= \left\langle \frac{p_n}{p_r+p_n+p_c} \right\rangle \\
 &\approx \frac{f_n \phi_r-f_r \phi_n+f_r \phi_n {e}^{k_t \phi_r t}}{f_c \phi_r+f_n \phi_r-f_r \phi_c-f_r \phi_n+f_r {e}^{k_t \phi_r t}}+\frac{F_M(k_t,\phi_r,\phi_n,\phi_c,f_r,f_n,f_c,t)}{N} \\
 &= \frac{\langle p_n \rangle}{\langle p_r+p_n+p_c \rangle} + \frac{F_M(k_t,\phi_r,\phi_n,\phi_c,f_r,f_n,f_c,t)}{N}.
\end{aligned}
\end{equation}
As the system size $N\to\infty$, the terms that depend on $N$ in (\ref{ENCyTaylor}) go to zero, and we have
\begin{equation}
 \langle \Phi_{NCell}(t) \rangle \to \frac{f_n \phi_r-f_r \phi_n+f_r \phi_n {e}^{k_t \phi_r t}}{f_c \phi_r+f_n \phi_r-f_r \phi_c-f_r \phi_n+f_r {e}^{k_t \phi_r t}} = \frac{\langle p_n(t) \rangle}{\langle p_r(t) \rangle+\langle p_n(t) \rangle+\langle p_c(t) \rangle},\quad N\to\infty.
\end{equation}
In other words, as the system size increase, the mean of the ratio approaches the ratio of the mean (Fig. \ref{NCeSystemSizeComparisonTaylor}).

The expressions for the variances of $\Phi_{NCell}$ has the form
\begin{equation}
\mathrm{Var}(\Phi_{NCell}) \approx \frac{\langle p_n \rangle^2}{\langle p_r+p_n+p_c \rangle^2}\cdot\frac{F_V(k_t,\phi_r,\phi_n,\phi_c,f_r,f_n,f_c,t)}{N},
\end{equation}
so $\mathrm{Var}(\Phi_{NCell})\to 0$, $N\to\infty$. That is, fluctuations of the $\Phi_{NCell}$ ratio become negligible for sufficiently large values of the system size $N$ (Fig. \ref{NCeVarianceComparison}).

\subsection{N/C ratio fluctuation and cell division}\label{Division}

We next study the effects of cell division on the homeostasis of the $\Phi_{NCell}$ ratio. When cells divide, molecules and organelles are randomly segregated and distributed into the two daughter cells. A simple model of this random partitioning assumes that each molecule or organelle has a 50/50 chance to enter either of the two daughter cells, which then leads to binomial distributions. In reality, the segregation mechanisms are more complex, which would give rise to various degrees of randomness in the resulting distribution \cite{huh2011random}.

\begin{figure}[h]
 \includegraphics[width=0.4\textwidth]{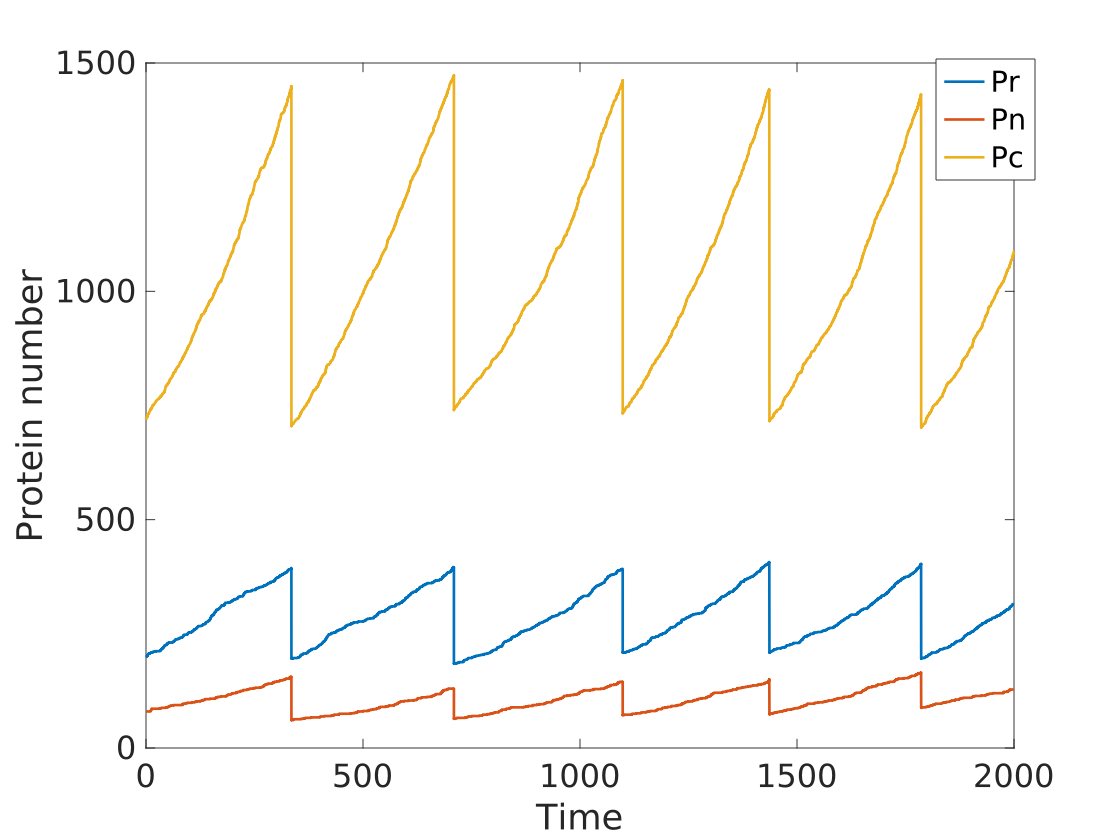}
 \caption{Sample protein number curves with 5 divisions. System size $N=1000$.}
 \label{Division_protein_count}
\end{figure}

\begin{figure}[h]
 \includegraphics[width=0.8\textwidth]{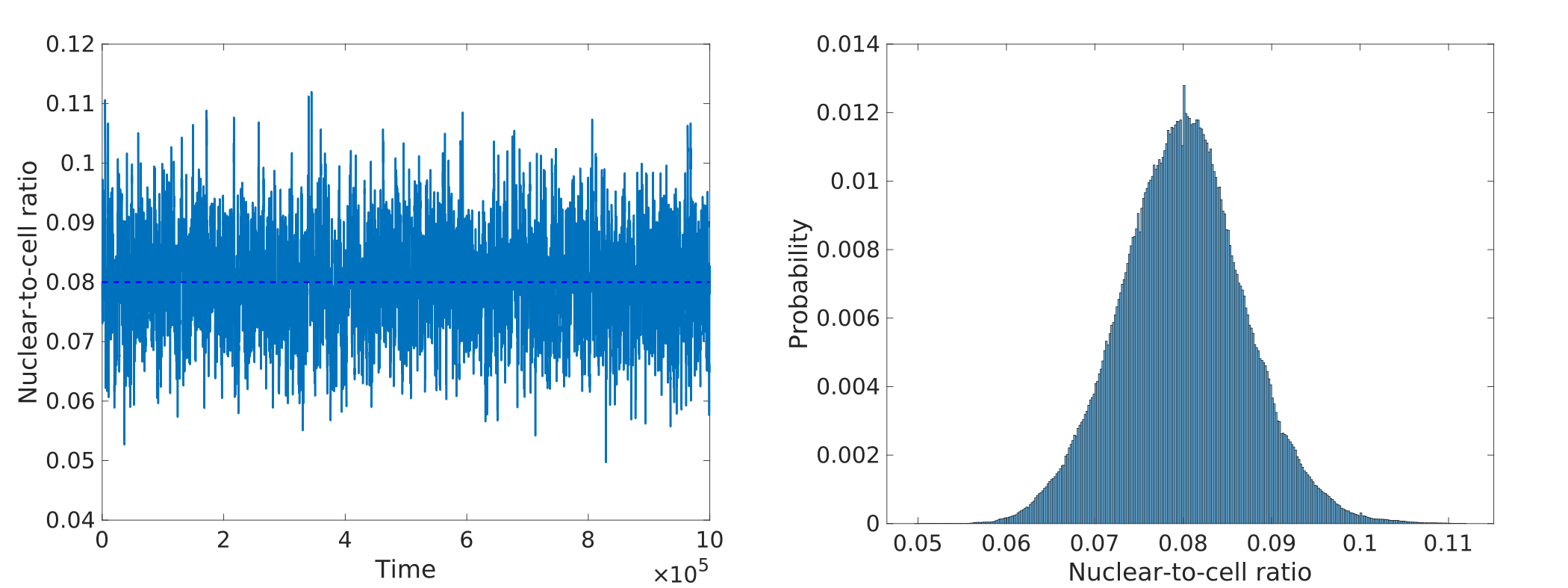}
 \caption{Left: sample $\Phi_{NCell}$ ratio curve with 2870 divisions. Right: corresponding histogram of the $\Phi_{NCell}$ ratio. System size $N=1000$.}
 \label{Division_NCe}
\end{figure}

Incorporating this simple model of cell division into the translation model is straightforward when using Gillespie algorithm for simulation. For instance, we may set a threshold for cell division by keeping track of an ``initiator'' protein corresponding to the DNA replication process \cite{lin2018homeostasis}. Alternatively, we may specify a \textit{protein doubling} division criterion by simply assuming that division occurs when the total number of proteins reaches twice the initial number. That is, the cell will divide whenever $p_r+p_n+p_c\geqslant 2(p_{r,0}+p_{n,0}+p_{c,0})$. After division, we track one of the daughter cells and set new initial values of protein numbers from binomial distributions based on the protein numbers at division. Fig. \ref{Division_protein_count} shows sample curves of protein numbers from the \textit{protein doubling} model with 5 divisions, and Fig. \ref{Division_NCe} shows a sample $\Phi_{NCell}$ ratio curve and the corresponding histogram with 2870 divisions. We conjecture that as the number of divisions becomes large, the probability distribution of the $\Phi_{NCell}$ ratio will approach a stable distribution, but further study is needed to reach a definite conclusion.

\begin{figure}[h]
 \includegraphics[width=0.4\textwidth]{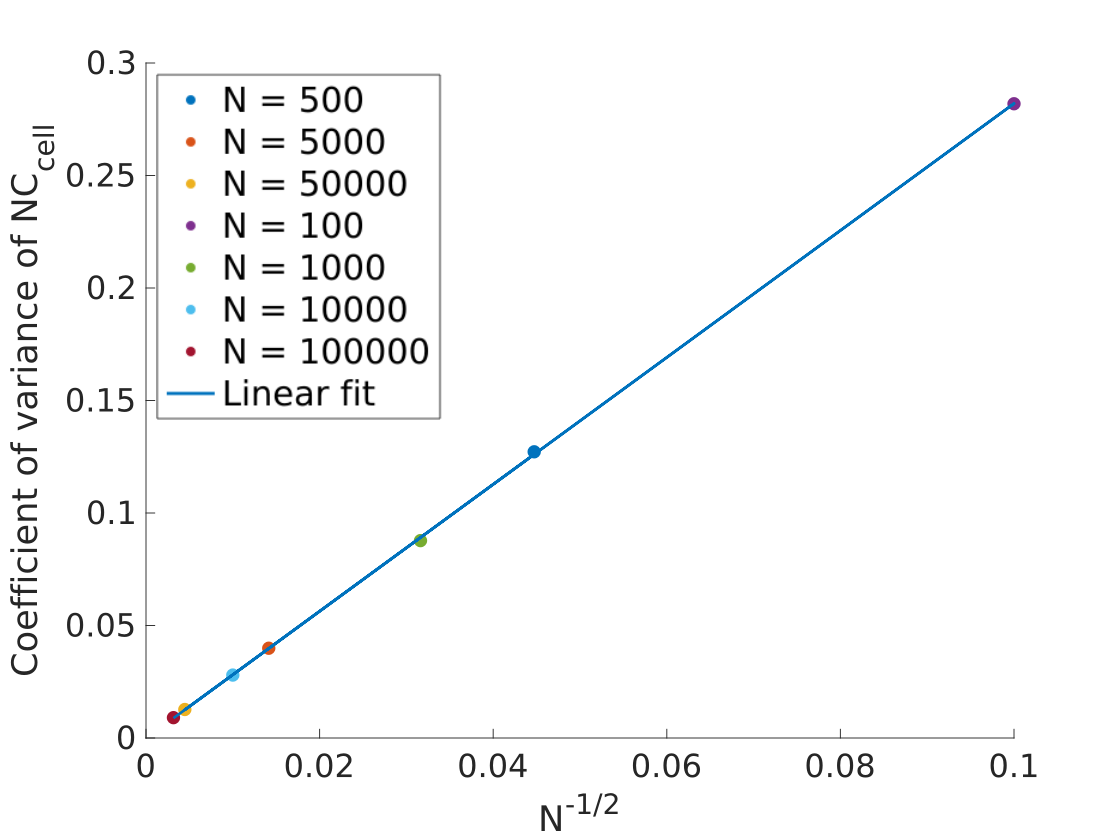}
 \caption{Plot of the coefficient of variance of the $\Phi_{NCell}$ ratio as a function of the inverse square root of the system size variable $N$.}
 \label{NCRatioDivCVLinearRegression1}
\end{figure}
We observe that as the system size variable $N$ becomes large, the relative fluctuations in the $\Phi_{NCell}$ ratio becomes smaller, even in the presence of extrinsic noises from cell division. We next examine $\mathrm{CV}(\Phi_{NCell})$, the coefficient of variance of the $\Phi_{NCell}$ ratio, as a function of $N^{-\frac12}$, where 
\begin{equation*}
 \mathrm{CV}(\Phi_{NCell}) = \frac{\mathrm{SD}(\Phi_{NCell})}{\langle \Phi_{NCell} \rangle} = \frac{\sqrt{\mathrm{Var}(\Phi_{NCell})}}{\langle \Phi_{NCell} \rangle},
\end{equation*}
and the mean $\langle \Phi_{NCell} \rangle$ and variance $\mathrm{Var}(\Phi_{NCell})$ are taken from all the data points in one SSA sample path, as is shown in Fig. \ref{Division_NCe}. The result (Fig. \ref{NCRatioDivCVLinearRegression1}) indeed shows that $\mathrm{CV}(\Phi_{NCell})$ decreases as $N$ increases; moreover, the relationship between $\mathrm{CV}(\Phi_{NCell})$ and $N^{-\frac12}$ is linear, a sign that the law of large number holds for this simple growth-division model. This means that fluctuations of the $\Phi_{NCell}$ ratio become negligible when the system size $N$ becomes large despite the extrinsic noises from cell division.

\section{Discussion}

Our simplified protein translation model, built upon our previous deterministic growth model \cite{lemiere2022control}, successfully demonstrates that the N/C ratio homeostasis is maintained when the stochasticity in cell growth is taken into account. This result holds in the case of unrestricted growth, which is clearly not biologically realistic due to resource limitations, as well as for a simple model of cell division, which we briefly explored in \ref{Division}.

It is estimated that the order of magnitude of the typical concentration of proteins in cells is $10^{-1}$mM$\sim$1mM \cite{biswas2023conserved,rollin2022cell}. Assuming that the volumes of cells are on the order of $10^2$\textmu m$^3$ \cite{lemiere2022control,wang2017thermodynamic}, the total number of proteins in a cell is expected to be on the order of $10^7\sim 10^8$. Our Taylor expansion approximation of the N/C ratio then shows that for this order of magnitude of proteins, the fluctuations in the N/C ratio are expected to be negligible. However, we have observed in our previous experiments that the N/C ratio does fluctuate in time (Figure 7C in \cite{lemiere2022control}), although it is unclear to what extent these fluctuations are caused by measurement errors. Below we will discuss a number of possible sources of fluctuations in the N/C ratio which are not included in the current simplified model, and which may be the topic of future research.

It was shown in a previous stochastic growth model that the dynamics of mRNAs in the transcription process may be a major source of growth rate fluctuations \cite{thomas2018sources}. However, in order to make the CME analytically solvable, we have removed transcription from the model entirely by assuming the translation rates are directly proportional to the gene fractions, thereby omitting this source of fluctuations.

In \cite{thomas2018sources}, transcription is modeled by a zeroth order reaction; whereas translation is modeled by a Michaelis-Menten kinetics for the ribosomes. Combining this model and the model in \cite{lin2018homeostasis} which our current model is based on yields a transcription-translation model that assumes Michaelis-Menten kinetics for both the RNAPs and the ribosomes:
\begin{equation}
 P_p+G_i\xleftrightarrow[k_{g,i}^-]{k_{g,i}^+}C_{g,i},\quad C_{g,i}\xrightarrow{k_0}P_p+G_i+M_i,\quad i = p,r,n,c \quad\mbox{(transcription);}
\end{equation} 
\begin{equation}
 P_r+M_i\xleftrightarrow[k_{m,i}^-]{k_{m,i}^+}C_{m,i},\quad C_{m,i}\xrightarrow{k_t}P_r+M_i+P_i,\quad i = p,r,n,c \quad\mbox{(translation);}
\end{equation}
\begin{equation}
 M_i\xrightarrow{\frac{1}{\tau}}\emptyset,\quad i = p,r,n,c \quad\mbox{(mRNA degredation).}
\end{equation}
It can be shown \cite{jachimowski1964stochastic} that if 
\begin{enumerate}[label=(\roman *)]
 \item the binding rate constants are identical for all genes, i.e. 
 \begin{equation*}
  k_{g,i}^+ = k_g^+,\,k_{g,i}^- = k_g^-,\,k_{m,i}^+ = k_m^+,\,k_{m,i}^- = k_m^-;
 \end{equation*}
 \item the forward rate constants are large enough, i.e. $k_g^+,k_m^+\to\infty$,
\end{enumerate}
 then at steady state, the average numbers of the enzyme-substrate complexes $c_{g,i}$ and $c_{m,i}$ will be
\begin{equation}
 \langle c_{g,i} \rangle = \frac{\langle g_i \rangle \langle p_{p,total} \rangle}{\sum \langle g_i \rangle}\quad\mbox{and}\quad \langle c_{m,i} \rangle = \frac{\langle m_i \rangle \langle p_{r,total} \rangle }{\sum \langle m_i \rangle},
\end{equation}
where $p_{p,total}$ and $p_{r,total}$ are the total number of RNAPs and ribosomes, respectively. In this case, the Michaelis-Menten model is essentially equivalent to the model in \cite{lin2018homeostasis}. Note, however, that in reality the above two assumptions may not hold. For example, in \cite{balakrishnan2022principles} the authors show that in \textit{E. coli} the promoter on rates, which correspond to $k_{g,i}^+k_0$ in our Michaelis-Menten model, span more than three orders of magnitudes across genes, and are modulated by the cells to regulate gene expression. Therefore, it is unlikely the Michaelis-Menten model would be equivalent to the  model in \cite{lin2018homeostasis}. 

Besides including the fluctuations arising from transcription, this Michaelis-Menten model may reveal other possible sources of fluctuations in the N/C ratio as well. According to \cite{holehouse2020stochastic}, the marginal distributions in reactions described by the Michaelis-Menten mechanism may become bimodal at intermediate times. In such cases, the coefficients of variation may no longer be of order $N^{-\frac12}$ and fluctuations in protein numbers may still be prominent even if the system size gets large.

In our current model, the N/C ratio is calculated by simply calculating the ratio of nuclear and cytoplasmic protein numbers. In reality, nuclear proteins are translated in the cytoplasm and need to diffuse through a heterogeneous environment \cite{chae2024beyond} before being imported into the nucleus by the nucleocytoplasmic transport process \cite{wang2017thermodynamic}. It is shown in \cite{leech2022mathematical} and \cite{wu2022correlation} that nucleocytoplasmic transport plays an important role in maintaining the N/C ratio. The nucleocytoplasmic transport has been studied extensively using both deterministic models (e.g. \cite{kim2013simple}, \cite{wang2017thermodynamic}) and stochastic models (e.g. \cite{moussavi2011brownian}). However, whether or not the diffusion and transport of nuclear proteins has significant contribution to the fluctuations in the N/C ratio is still unknown.

The random partioning of nuclear and cytoplasmic proteins at cell division may be yet another source of fluctuations in the N/C ratio for cells that are actively dividing. We have shown that in a simple growth-division model where the partitioning of molecules follow a binomial distribution with probability $1/2$, the fluctuations in the N/C ratio become negligible as the system size becomes sufficiently large. However, more complicated partioning mechanisms may lead to significant fluctuations in the N/C ratio. In particular, if the proteins are not independently partitioned at cell division but are partitioned into large clusters instead \cite{huh2011random}, then the induced fluctuation may be non-negligible even if the system size is large.

There are other potential sources of fluctuations in the $\Phi_{NCell}$ ratio. For example, in \cite{cadart2022volume} the authors hypothesized that fluctuations in cell volume may originate from biophysical sources, such as membrane tension and osmotic pressure. Admittedly, biochemical and biophysical processes in live cells are far more complicated than what our simplified protein translation model has encompassed. We believe that this simple model provides a foundation for future work on more realistic stochastic models of N/C ratio determination mechanisms.

\section*{Acknowledgments}

We acknowledge useful discussions with Sam Isaacson and Peter Thomas. We acknowledge funding from NSF grant MCB-2213583.

\printbibliography
\end{document}